# *EndWatch*: A Practical Method for Detecting Non-Termination in Real-World Software

Yao Zhang*, Xiaofei Xie†§, Yi Li‡, Sen Chen*§, Cen Zhang‡, Xiaohong Li*§

*College of Intelligence and Computing, Tianjin University
†School of Computing and Information Systems, Singapore Management University
‡School of Computer Science and Engineering, Nanyang Technological University

*Abstract*—Detecting non-termination is crucial for ensuring program correctness and security, such as preventing denial-of-service attacks. While termination analysis has been studied for many years, existing methods have limited scalability and are only effective on small programs. To address this issue, we propose a practical termination checking technique, called *EndWatch*, for detecting non-termination caused by infinite loops through testing. Specifically, we introduce two methods to generate non-termination oracles based on checking state revisits, *i.e.*, if the program returns to a previously visited state at the same program location, it does not terminate. The non-termination oracles can be incorporated into testing tools (*e.g.*, AFL used in this paper) to detect non-termination in large programs. For linear loops, we perform symbolic execution on individual loops to infer *State Revisit Conditions* (SRCs) and instrument SRCs into target loops. For non-linear loops, we instrument target loops for checking concrete state revisits during execution. We evaluated *EndWatch* on standard benchmarks with small-sized programs and real-world projects with large-sized programs. The evaluation results show that *EndWatch* is more effective than the state-of-the-art tools on standard benchmarks (detecting 87% of non-terminating programs while the best baseline detects only 67%), and useful in detecting non-termination in real-world projects (detecting 90% of known non-termination CVEs and 4 unknown bugs).

## I. INTRODUCTION

Ensuring correctness is an essential part of software quality assurance. A program is considered *partially correct* if it produces correct results whenever it terminates. Moreover, it is considered totally correct if it terminates and produces correct results [1]. Therefore, ensuring the total correctness of a program requires proving both its partial correctness and termination.

To ensure program correctness, the study of program termination has a long history. The basic idea of proving non-termination of a program is to find a *recurrent set* in which the program stays indefinitely [2]. Many verification tools such as CPAChecker [3], UAutomizer [4], and AProVE [5] have been developed to detect non-termination. However, their scalability is limited, making them ineffective in detecting non-termination in large programs, particularly for real-world software projects. The main challenge in program termination analysis is the undecidability of finding recurrent sets, which becomes particularly expensive for large programs with numerous loops, complex program features, and long execution traces. A recent study [6] demonstrated that the state-of-the-art loop-analysis tools fail to handle real-world programs directly. These tools did not perform well even on substantially simplified real-world programs due to presence of complex features, such as floating point numbers, arrays, bit shifts, and pointer arithmetics. Furthermore, the study revealed that real-world non-termination bugs are caused by not only common logical errors (*e.g.*, mistakes in loop conditions or loop variables), but also misuses of low-level program features (*e.g.*, overflows and variable type casts). These findings motivate us to develop a technique which is able to detect non-termination in large real-world programs.

In contrast to software verification, testing is scalable and works well on large real-world programs. There have been numerous testing techniques [7]–[9] proposed to detect bugs by identifying test inputs that lead to erroneous program states. However, testing cannot be used directly in detecting non-termination bugs, due to the lack of test oracles. Unlike other bugs (*e.g.*, functional bugs and memory corruptions) which can be detected by designing specific test oracles (*e.g.*, program crashes, assertions, and address sanitizers [10]), the violation witnesses of a non-termination bug are infinite traces that cannot be captured by a traditional test oracle. In particular, fuzz testing tools, such as AFL [11], have been shown to be effective in finding bugs and security vulnerabilities in real-world programs, such as buffer overflow, use-after-free bugs, and other memory corruption issues. However, they still do not possess the ability to detect non-termination bugs. Similar to existing performance issue detection methods [12]–[18], AFL can generate *hang tests*. However, it remains unclear whether these hangs are due to non-termination bugs or performance degradation, which are distinct issues. The latter is often considered a non-functional problem rather than a correctness issue. Therefore, despite significant progress in theory and tool development over the years, proving non-termination of real-world software remains an open problem due to the limited scalability in verification methods and the lack of oracles in testing methods.

To address this gap, we propose a practical testing method *EndWatch* for detecting non-termination in real-world software projects. Our approach involves generating non-termination oracles by detecting *state revisits* during program executions, where the intuition is that if the program returns to a previously visited state, it does not terminate. To tackle scalability issues, we adopt a divide-and-conquer strategy that generates non-termination oracles for each loop[1] individually, rather than the entire program. Specifically, for linear loops, we propose a symbolic execution-based method that infers the State Revisit Condition (SRC), which is the weakest precondition that two *symbolic* states visited in a loop execution are equivalent. If the corresponding SRC is satisfied during testing, we discover a state revisit that can serve as a non-termination witness. For other loops where *EndWatch* cannot infer the SRC, we instrument the loops with revisit monitors to detect revisits of *concrete* states during program executions.

We designed two experiments to evaluate the effectiveness of *EndWatch*. To compare *EndWatch* with the state of the art, we selected three non-termination benchmarks including SV-COMP [19], TermComp [20], and OSS_Bench [6], where OSS_Bench [6] contains simplified programs of real-world non-termination bugs. The results show that *EndWatch* correctly detects 87% of the non-terminating programs while the best baseline (*i.e.*, UAutomizer) only correctly handles 67%. To evaluate the usefulness of *EndWatch* in real-world software projects, we collected 12 projects that contain a total of 20 known CVEs related to non-termination. *EndWatch* successfully identified 90% of the CVEs, except the two where we were unable to obtain the source code. We further ran *EndWatch* on real-world projects and 4 unknown bugs were detected, where two of them have been confirmed and fixed by the developers.

To summarize, we made the following contributions:
- We proposed a practical testing technique to detect non-termination in large real-world programs.
- We proposed two methods to generate non-termination oracles needed for discovering non-termination bugs: inferring state revisit conditions for linear loops and monitoring concrete state revisits for non-linear loops.
- We conducted experiments to evaluate the effectiveness of *EndWatch* on standard benchmarks and its usefulness on real-world projects. In particular, *EndWatch* discovered 4 new bugs. All detailed results and source code can be found at https://sites.google.com/view/endwatch/home.

## II. PRELIMINARY

In this section, we introduce definitions and preliminaries.

*Definition 1:* A control flow graph (CFG) $\mathcal{G}$ of a loop is defined as a tuple $\mathcal{G} = (X, B, E, B_h, B_e)$:
- $X$ denotes the variables in the loop. A state $s$ of the loop is a valuation of the variables from $X$.
- $B$ is a set of basic blocks. Each basic block $b \in B$ contains a sequence of statements that construct a transition relation $\rho_b$ over the variables $X \bigcup X'$, where $X'$ denote the updated values of the variables $X$ after the execution of the basic block. We define $(s, s') \models \rho_b$ if the valuation $s$ from $X$ and the valuation $s'$ from $X'$ satisfy the constraint $\rho_b(X, X')$.
- $E \subseteq B \times C \times B$ is a set of edges representing the transitions between blocks, where $C$ is a set of conditions. An edge $b_i \xrightarrow{c} b_j$ is feasible if $c \in C$ can be satisfied.
- $B_h$ is a set of header blocks. Especially, a loop can have multiple header blocks if it is a nested loop.
- $B_e$ is a set of exit blocks from which the loop can terminate.

---
[1]In this paper, we mainly focus on non-termination caused by infinite loops.

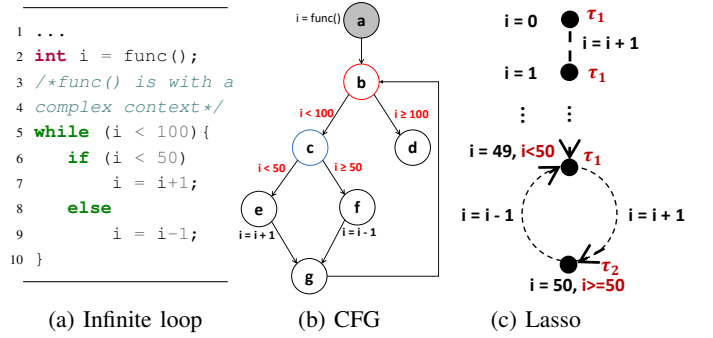

Fig. 1: An example of a loop with its CFG and lasso

Figure 1a shows an infinite loop, and Fig. 1b is the corresponding CFG. It has one variable, *i.e.*, $X = \{i\}$ and $X' = \{i'\}$. $B = (a, b, c, d, e, f, g)$ is the set of basic blocks and $i = i+1$ is a statement in the block $e$. The transition relation $\rho_e$ can be written as $i' = i + 1$. The two states $s : i = 1$ and $s' : i' = 2$ satisfy this relation, *i.e.*, $(i = 1, i' = 2) \models \rho_e$. The edge $c \xrightarrow{i<50} e$ is feasible if and only if $i < 50$ is satisfied. The basic block $d$ is an exit block.

*Definition 2:* A *loop path* $\tau = b_1 \xrightarrow{c_1} \ldots \xrightarrow{c_{n-1}} b_n$ is a sequence of blocks beginning and ending with a header block or an exit block, i.e., $b_1 \in B_h \bigwedge b_n \in B_h \cup B_e$:
- The transition relation of the path can be represented by $\rho_\tau = \rho_{b_1} \circ \ldots \circ \rho_{b_n}$, where $\circ$ is the relational composition operator [2].
- The *loop path condition (LPC)* $\theta_\tau$ is defined as $wp(\rho_{b_1}, c_1) \bigwedge wp(\rho_{b_1} \circ \rho_{b_2}, c_2) \bigwedge \ldots \bigwedge wp(\rho_{b_1} \circ \ldots \circ \rho_{b_{n-1}}, c_{n-1})$, where $wp(\rho, c)$ calculates the weakest precondition of $c$ over the relation $\rho$. The loop path $\tau$ is feasible if the LPC $\theta_\tau$ can be satisfied.

The CFG has three loop paths, *i.e.*, $\tau_1 = b \xrightarrow{i<100} c \xrightarrow{i<50} e \xrightarrow{true} g \xrightarrow{true} b$, $\tau_2 = b \xrightarrow{i<100} c \xrightarrow{i\geq 50} f \xrightarrow{true} g \xrightarrow{true} b$, and $\tau_3 = b \xrightarrow{i\geq 100} d$. The transition relation for $\tau_1$ is $\rho_{\tau_1} = \rho_b \circ \rho_c \circ \rho_e \circ \rho_g$, *i.e.*, $i' = i + 1$. The path condition for $\tau_1$ is $\theta_{\tau_1} = \{i < 100\} \wedge \{i < 50\}$.

*Definition 3:* A loop execution can be viewed as a lasso [21] which consists of two sequences of transitions, referred to as *stem* and *loop*:

$$lasso = \tau_0 \xrightarrow{stem} \tau \xrightarrow{loop} \tau$$

where the *stem* is a finite sequence starting from $\tau_0$ and the *loop* is a finite sequence that starts and ends at $\tau$.

Note that a loop in the program can have different executions based on different inputs, representing different lassos. Intuitively, an execution is non-terminating if the loop sequence of the lasso can be unrolled infinitely.

Figure 1c shows a lasso of the loop in Fig. 1a. Given a variable $i = 0$, it first increases to 49 on $\tau_1$, which can be seen as the stem. Then once $i$ reaches 50, it will cyclically transits between $\tau_2$ and $\tau_1$, falling into a loop of the lasso (*i.e.*, $\tau_1 \longrightarrow \tau_2 \longrightarrow \tau_1$).

*Definition 4:* For a transition relation $\rho$, we say that $\mathcal{R}_\rho(X)$ is a *Recurrent Set*, if for each state $s \in \mathcal{R}_\rho(X)$, there exists a state $s'$ such that $s' \in \mathcal{R}_\rho(X) \bigwedge (s, s') \models \rho$.

*Proposition 1:* A lasso $\tau_0 \xrightarrow{stem} \tau \xrightarrow{loop} \tau$ is non-terminating if there exists $s_0, s_1$ and an recurrent set $\mathcal{R}_{\rho_{loop}}(X)$ for $\rho_{loop}$ such that $(s_0, s_1) \models \rho_{stem} \bigwedge s_1 \in \mathcal{R}_{\rho_{loop}}(X)$, where $\rho_{stem}$ and $\rho_{loop}$ are the transition relations of the sequences *stem* and *loop*, respectively.

Note that $(s_0, s_1) \models \rho_{stem} \bigwedge s_1 \in \mathcal{R}_{\rho_{loop}}(X)$ represents that the initial state $s_0$ can reach the recurrent set.

In Fig. 1c, the loop begins with $i = 0$ and increases $i$ to 50, and it has a recurrent set $\{i = 49, i = 50\}$. The initial state $s_0$ is $i = 0$ and there exists a state $s_1$, *i.e.*, $i = 49$ that reaches the recurrent set. Therefore, the loop is non-terminating.

## III. OVERVIEW

### A. Problem Definition

In general, the execution of a program can be viewed as a concatenation of multiple lassos,

$$\langle \tau_0 \xrightarrow{stem_1} \tau_1 \xrightarrow{loop_1} \tau_1 \rangle \ldots \langle \tau_{n-1} \xrightarrow{stem_n} \tau_n \xrightarrow{loop_n} \tau_n \rangle.$$

The number of lassos depends on the number of loops in the execution. Hence, larger executions with a greater number of loops are likely to have more lassos, resulting in more complexity for termination analysis.

*Definition 5 (Program Non-Termination):* A program is non-terminating if there exists an initial state $s_0$ that leads to a recurrent-set belonging to one of the lassos, *i.e.*, $\langle \tau_0 \xrightarrow{stem_1} \tau_1 \xrightarrow{loop_1} \tau_1 \rangle \ldots \langle \tau_{n-1} \xrightarrow{stem_n} \tau_n \xrightarrow{loop_n} \tau_n \rangle$, where the state $s_0$ is the input to the program.

To prove non-termination of a program, we need to find inputs (*i.e.*, initial state $s_0$) and a *lasso* of the program such that the input can reach a recurrent-set inferred from the *lasso*.

*Differences from detecting performance issues.* It is worth noting that some existing works [12]–[18] focus on detecting performance issues, such as inefficient executions. This paper focuses on detecting and proving non-termination instead, which differs at the problem definition. While detecting performance issues can be accomplished by checking execution time, proving non-termination is more difficult and requires showing that a program *never* terminates. A long-running execution may indicate a performance issue, but it is not necessarily an indicator for non-termination.

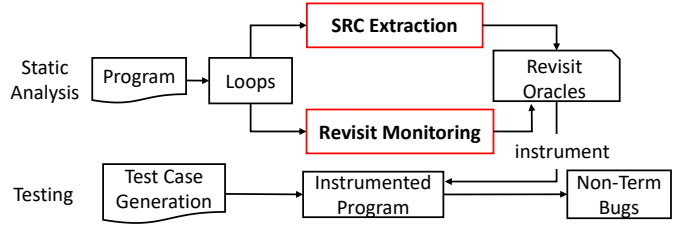

Fig. 2: Overview of our work

### B. Overview of EndWatch

**Challenges.** There are two main challenges that the state-of-the-art tools face when detecting non-termination in real-world programs. ① Real-world programs are often large and contain many loops, making it difficult to statically determine the context of deep loops, which is required for recurrent-set analysis. As shown in Definition 5, when $n$ is large, calculating the recurrent-set for the last lasso $\tau_n \xrightarrow{loop_n} \tau_n$ becomes challenging, as its context depends on the postcondition of the previous loops and function calls. For example, the loop in Fig. 1a requires the analysis of function *func* and others. ② Inferring recurrent-sets is undecidable and cannot be fully automated, especially for complex loops that involve pointers, arrays, and custom data structures.

**Our Solution.** To address these challenges, we propose a practical and scalable method called *EndWatch*, as shown in Fig. 2. To tackle challenge ①, we adopt a divide-and-conquer strategy that analyzes each loop individually instead of the entire program. We extract all loops in a program and calculate non-termination oracles for each loop. As a testing approach, *EndWatch* has access to runtime information about the program and can easily determine the precise context of each individual loop. To address challenge ②, we propose to sacrifice on generality, and design more practical and tractable non-termination oracles. We employ two strategies for generating non-termination oracles: *i.e.*, State Revisit Condition (SRC) extraction and Revisit Monitoring. Specifically, for linear loops, we perform symbolic execution to infer SRCs, so that the loop is non-terminating if an SRC is satisfied by the test input. For other loops whose SRCs cannot be determined, we implement runtime monitors to check for concrete state revisits by comparing the current state with the previously visited states. Of course, revisit monitoring is more resource-intensive, as it requires recording and comparing concrete program states. Both SRCs and revisit monitors are instrumented into the program as non-termination oracles during testing.

Figure 3 illustrates the two types of oracles generated by *EndWatch*. The condition calculated by *SRC extraction* is $i = 49$, which serves as a non-termination oracle for the loop: if there exists an input that makes $i$ equal to 49 before entering the lasso, a non-termination is detected. The *revisit monitoring* analysis involves dynamically recording previously visited concrete states, such as $\{20, 40, 50\}$ for $i$. These states are used to compare with subsequent states during loop execution.

**Algorithm 1:** InferSRC

**Input :** $\tau_0$: An initial path of the loop;
$Q$: A set of all paths ;
**Output:** $C$: State Revisit Condition;

1 Create an initial symbolic state $s_0$ ;
2 $s_0.TC = \theta_{\tau_0}$;
3 $s_0.trace \leftarrow [\tau_0]$;
4 $worklist \leftarrow \{s_0\}$;
5 **while** $worklist \neq \emptyset$ **do**
6    $s \leftarrow SELECT(worklist)$;
7    $\tau \leftarrow s.trace[-1]$;
8    **foreach** $\tau' \in Q$ **do**
9      **if** $head(\tau') = tail(\tau)$ **then**
10        $s' \leftarrow FORK(s)$;
11        Introduce a variable $k$ to represent the $k$ iterations of $\tau$;
12        **if** $INDUCTIVE(\tau)$ **then**
13           $\phi_k \leftarrow \theta_\tau[X/X'_{k-1}] \wedge \theta_{\tau'}[X/X'_k]$ ;
           /* Constraints on $k$ */
14        **else**
15           $\phi_k \leftarrow k = 1$ ;   /* No summary for non-inductive paths */
16        $s'.TC = s'.TC \wedge \theta_{\tau'}[X/X'_k] \wedge \rho_\tau^k \wedge \phi_k$ ;
       /* Perform path summarization with $k$ */
17        **if** $FEASIBLE(s'.TC)$ **then**
18           **if** $VISITED((\tau, \tau'), s')$ **then**
19              **continue** ;   /* State pruning */
20           $s'.trace.append(\tau')$;
21           $worklist \leftarrow worklist \cup \{s'\}$;
22           **if** $IS\_LASSO(s')$ **then**
23              Let $s'.trace = \tau_0 \xrightarrow{stem} \tau' \xrightarrow{loop} \tau'$;
24              $c \leftarrow INFER(s'.trace)$ ; /* Infer SRC */
25              $C \leftarrow C \cup \{c\}$;

26 **return** $C$;

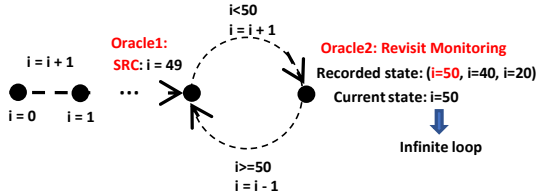

Fig. 3: Two types of oracles for the lasso in Fig. 1

For example, when a matching state of 50 is identified, a non-termination is detected.

## IV. METHODOLOGY

### A. State Revisit

State revisit in a lasso implies the existence of a recurrent-set for the lasso. Therefore, a loop does not terminate if we can find a state revisit, which is formalized as follows:

*Proposition 2:* A lasso $\tau_0 \xrightarrow{stem} \tau \xrightarrow{loop} \tau$ is non-terminating if there exists $s_0, s_1$ such that $(s_0, s_1) \models \rho_{stem} \bigwedge (s_1, s_1) \models \rho_{loop}$, where $\rho_{stem}$ and $\rho_{loop}$ are the transition relations of the sequences *stem* and *loop*, respectively.

*Proof:* Assume there is no non-deterministic operation in $\rho_{loop}$, if $(s_1, s_1) \models \rho_{loop}$, then there exists one reachable recurrent-set $\{s_1\}$, which serves as a proof of non-termination based on Proposition 1.

The problem of detecting non-termination is now reduced to finding state revisits for a given loop. We propose two types of oracles to detect state revisits: *symbolic state revisits* (Section IV-B) and *concrete state revisits* (Section IV-C). For symbolic state revisits, we perform symbolic execution to identify symbolic states that occur at the same location of the loop (*e.g.*, $X$ and $X'$). We then calculate the weakest State Revisit Condition (SRC) that makes the symbolic states equal (*i.e.*, $X = X'$). The SRC is then instrumented into the loop as the non-termination oracle. For concrete state revisits, we directly instrument the loop by recording the previously visited states and comparing them with the current state (*e.g.*, $s = s'$). During testing, a non-terminating loop is discovered if a state revisit occurs. A detailed comparison of the two methods will be discussed in Section IV-C.

### B. State Revisit Condition Extraction

Algorithm 1 outlines the process of identifying lassos within a loop and calculating their SRCs. The basic idea is to iterate the loop and compare the symbolic states. A major challenge in this approach is the possibility of encountering an infinite number of lassos during the symbolic execution (*i.e.*, state explosion). To address this challenge, we adopt two strategies: path summarization and state pruning, which can alleviate the state explosion problem. Path summarization calculates the effect of a path after any $k$ iterations, while state pruning limits the number of iterations.

Algorithm 1 takes as input a loop with an initial path $\tau_0$ and a set of paths $Q$. The algorithm starts by initializing a symbolic state $s_0$, which includes various properties such as the current trace[2] executed and its corresponding trace condition $s_0.TC$ (Lines 2-3). A standard worklist-based symbolic execution (as described in [22]) is then executed until the worklist is empty (Lines 5-25).

At each iteration, a symbolic state $s$ is selected and removed from the worklist by the $SELECT$ function (Lines 6-7). The function $SELECT$ can be implemented following various selection strategies, and in this paper, we always select the first state from the worklist. The algorithm then iterates through the path list to select the next path (Lines 8-25). The transition from $\tau$ to $\tau'$ requires that the last basic block of $\tau$ matches the first basic block of $\tau'$ (Line 9).

**Path Summarization (Lines 11-16).** The algorithm executes the current path $\tau$ symbolically and transits to the next path $\tau'$. To reduce the number of times that symbolic execution traverses a path, we perform path summarization for inductive paths by INDUCTIVE function (Lines 12-13), as described in [23]. This allows us to obtain the result of executing path $\tau$ any $k$ times in a single iteration, thereby reducing the overall time. For instance, if the variable is increased by a constant value in each execution of the path (*e.g.*, $x := x + 1$), then we can obtain the result after $k$ executions, *i.e.*, $x + k$. To implement this, we introduce a variable $k > 0$ to represent that the loop will execute $\tau'$ after $\tau$ has executed $k$ times. This implies the constraint that after $k - 1$ iterations of $\tau$, the path condition of $\tau$ is satisfied, and after $k$ iterations of $\tau$, the path condition of $\tau'$ is satisfied (Line 13). The expression $\theta_{\tau'}[X/X'_k]$ represents the results after $k$ executions of $\tau$ (*i.e.*, $\rho_\tau^k$), with variables in $X$ substituted by the variables in $X'_k$. If $\tau$ is a non-inductive path that we fail to summarize, we perform a standard symbolic execution that does not involve path summary (*i.e.*, $k = 1$ in Line 15). It updates the current trace condition to reflect the transition from executing the path $\tau$ $k$ times to $\tau'$, and then checks the satisfiability of the updated trace condition. The trace condition is updated to be the conjunction of the current trace condition and the condition derived from executing $\tau$ $k$ times and transitioning to $\tau'$ (Line 16). If the updated trace condition is not feasible, the algorithm will skip to the next path $\tau'$.

---
[2]A "trace" is equivalent to the term *program path* used in the symbolic execution literature. We use the term "trace" to distinguish from *loop paths* used in this paper.

**State Pruning (Lines 18-19).** The algorithm implements the similar state pruning [22] strategy to prevent state explosion during symbolic execution, caused by the cyclic execution between multiple paths. It limits the number of iterations for each cycle of execution by only considering one cycle of execution at a time (Lines 18-19). It may miss some state revisit checking, but reduces the complexity of the symbolic execution and makes *EndWatch* more practical. If the state is not pruned, the algorithm creates a new symbolic state $s'$ for the next path $\tau'$ and adds it to the worklist (Lines 20-21), allowing the algorithm to continue its iteration.

**SRC Inference (Lines 22-31).** If a lasso is detected during the execution, meaning that the next path $\tau'$ exists in the previous execution, the algorithm infers a sufficient condition for the non-termination of the identified lasso (Lines 22-23). We calculate the SRC by ensuring that the values of the variables $X_{\tau'}$ remain unchanged after one execution of the cycle, *i.e.*, $X_{\tau'} = X'_{\tau'}$, where $X_{\tau'}$ and $X'_{\tau'}$ represent the variables before and after one execution of the cycle, respectively. $INFER$ aims to eliminate the variables $k_i$ from the condition $s'.TC \land X_{\tau'} = X'_{\tau'}$. Here, we use the Z3 theorem prover [24] to eliminate the existentially quantified variables, which is easier to do due to the inductiveness of the paths.

Consider the loop in Figure 1b, and there is a lasso $\tau_1 \to \tau_2 \to \tau_1$. Then we can calculate its SRC with two quantifiers $k_1$ and $k_2$, *i.e.*, $\{\exists k_1 \in \mathbb{Z}, \exists k_2 \in \mathbb{Z} | k_1 > 0 \land k_2 > 0 \land i < 50 \land i + k_1 - 1 < 50 \land i + k_1 \geq 50 \land i + k_1 - (k_2 - 1) \geq 50 \land i + k_1 - k_2 < 50 \land i + k_1 - k_2 = i\}$. After eliminating $k_1$ and $k_2$, we can get a simplified SRC $\{i \geq 49 \land i \leq 49\}$, *i.e.*, $\{i = 49\}$.

**Other Optimizations.** Additionally, we provide several optimization strategies to enhance Algorithm 1, including stuck path detection, non-inductive path summarization, and overflow handling. Due to space limit, we provide more details and examples about these strategies on our website [25].

Stuck paths are paths that cannot transit to any other paths, and they may not cause state revisits in a finite execution. If the loop enters a stuck path, it does not terminate. To detect stuck paths, we check whether the path condition $\theta_\tau$ can always be satisfied for any number of iterations using a universal quantifier $k$, as shown below:

$$StuckCon \leftarrow \forall k > 0 \cdot \theta_\tau[X/X'_k] \land \rho_\tau^k$$

If $StuckCon$ is satisfied, we use Z3 to eliminate the universal quantifier $k$, and the result is a stuck condition that can serve as an oracle for non-termination.

Non-inductive paths could also be summarized if we have prior knowledge about some special program constructs. For example, bit shifting is a non-linear operation, but in certain situations, it can result in a constant-valued variable or has an upper/lower bound. For instance, if a positive value is shifted right iteratively, it will eventually become 0. We can use these to calculate the path summary. Similarly, for function calls (*e.g.*, *read*), some APIs have specific return values (*e.g.*, 0 after reaching EOF) that can be used as the summary. We have implemented the strategies to handle some non-inductive paths.

Overflow occurs when a value exceeds the maximum or minimum value that can be represented by its data type. Existing techniques usually assume that overflow does not occur, which can lead to incorrect conclusions in real-world program execution. To handle overflow, we constrain the range of the value when calculating the path summary. For example, if the instruction x-- occurs in a loop path, the summary without considering overflow is $x' = x - k$. However, when we consider overflow, the summary becomes $x' = (x - k) \mod 2^{16}$ if $x$ is an unsigned 16-bit integer.

*C. Revisit Monitoring*

For loops that cannot be handled in Algorithm 1, we instrument them with state revisit monitors to record and compare the concrete states visited in loop executions. The method consists of two components: a state slicing process to determine which variables should be recorded for a state, and adaptive state recording strategy to compare the recorded states.

**State Slicing.** Choosing proper state abstractions is important for detecting state revisits. In other words, one needs to decide which variables to record as a part of the loop state: recording variables that do not affect loop termination could result in missing non-termination bugs due to mismatched states; conversely, missing variables that affect loop termination could lead to false positives caused by partial state matching.

We adopt the idea from [26] to slice variables that affect the termination of the loop. The process starts by selecting variables in loop conditions and then performing data dependency analysis and control dependency analysis to identify any other variables that may have an impact on the loop termination. Additionally, we conduct special analysis on the dependencies of global variables, function parameters/returns, and pointers through a points-to analysis to ensure that all relevant variables are monitored.

**Adaptive State Recording.** Determining the period of state recording is another challenge in revisit monitoring, because state revisit may not occur after every iteration. It could occur after a varying number of iterations, such as after each iteration or after thousands of iterations. Recording the state at each iteration is intuitive but impractical as it significantly increases the costs. To strike a balance between precision and efficiency, we aim to select an appropriate number of states for checking, rather than recording the state at every iteration.

Specifically, we propose an adaptive strategy to determine the period of state recording. Our strategy is based on an interval $\mathcal{I}$, where we record the states at $\mathcal{I}^{th}, 2 \times \mathcal{I}^{th}, \ldots, n \times \mathcal{I}^{th}$ iterations of the loop. This interval is dynamically updated based on the number of loop iterations:

$$\mathcal{I} = \mathcal{I}_0 \times \lceil \frac{\#iter}{\alpha} \rceil,$$

where $\mathcal{I}_0$ is the initial interval, $\#iter$ is the number of iterations, and $\alpha$ is a discount factor to control the interval value. In this work, we set $\mathcal{I}_0 = 100$ and $\alpha = 10,000$. This means that if the number of iterations is less than $10,000$, we record the states with a smaller interval ($100$) for finer-grained monitoring. However, if no state revisit is detected after a large number of iterations, we increase the interval to reduce the overhead of recording and comparing many states.

Finally, given a concrete state in the current iteration, if it can be matched with one of the recorded states, then the program does not terminate.

**SRC Extraction versus Revisit Monitoring.** The two methods complement with each other and work in synergy to detect non-termination bugs in real-world programs. The SRC extraction method is relatively light-weight for testing as it does not require recording too many states, and the inference of SRCs can be conducted offline, *i.e.*, before testing. Hence, it is suitable for linear programs (*e.g.*, benchmarks used by the existing tools). On the other hand, revisit monitoring tends to be more resource-intensive due to the need of recording states of running loops, especially when the state revisit cannot be detected within a small number of iterations. However, revisit monitoring tends to be more scalable than SRC extraction as it is able to handle non-linear loops.

### D. Instrumentation and Testing

*EndWatch* performs instrumentation to insert the oracles generated by SRC extraction and revisit monitoring. We use Clang to instrument the oracles into LLVM-IR code. For SRC, we insert it into the loop header as an asserted condition that reports non-termination. For revisit checking, our instrumentation consists of two parts, namely, the state recorder and the state checker. The state recorder is instrumented at the loop header to collect states, while the state checker is instrumented at the loop tail to compare the current state with the recorded states. To reduce the overhead of variable comparisons, *EndWatch* only records the hash value for all the sliced variables in the state.

With the instrumented code, we can then use testing tools to generate test cases and detect non-termination bugs based on the oracles. Note that the usage of testing tools is orthogonal to our method. In this paper, we used AFL, a widely used fuzzing tool to generate test cases.

### E. Discussions

*1) Soundness and Completeness:* As proven in Proposition 2, *EndWatch* is sound, meaning that if it finds a non-terminating test case, the program must not terminate. However, like other non-termination tools, *EndWatch* is not complete as non-termination proving is an undecidable problem. This means that *EndWatch* may miss some non-termination test cases. Theoretically, without considering scalability challenges, a recurrent-set based method would detect more bugs than *EndWatch* since *EndWatch* mainly considers state revisits, which is a special case of the recurrent-set.

*Limitation*. The primary limitation of *EndWatch* is its balance between practicality and generality. While *EndWatch* succeeds in achieving practicality, it scarifies the generality. This means that certain instances of non-termination can not be detected by *EndWatch*. Specifcially, the state pruning strategy mitigates state explosion issues, yet at the cost of potentially disregarding state revisits requiring more iterations. The adaptive state recording approach opt to capture only partial states, leading to commendable efficiency gains, but possibly overlooking certain instances of non-termination if the partial assessment is inadequate.

*2) Usage of EndWatch:* *EndWatch* offers two usage modes, namely, offline mode and online mode. For the online mode, *EndWatch* first instruments the program with oracles and then tests the instrumented programs. For the offline mode, we first generate test cases for the original programs without *EndWatch*. Then we collect the tests that do not terminate within a pre-defined threshold (*e.g.*, hangs in AFL's terminology). Although these slow test cases may not necessarily be non-terminating, they are more likely to be non-terminating. We can run *EndWatch* on them offline to confirm the non-terminating cases. The offline mode is more efficient as it only needs to analyze the slow loops instead of all the loops. But it relies entirely on the ability of the selected testing tool to detect potential non-terminating cases. On the other hand, the online mode is more expensive but can provide better feedback for the testing algorithm, such as non-termination guidance from the SRC and state revisit checking.

## V. EVALUATION

To evaluate the effectiveness of *EndWatch*, we aim to answer the following research questions:

- **RQ1**: How effective is *EndWatch* on existing benchmark programs compared with the state-of-the-art tools?
- **RQ2**: How effective is *EndWatch* on detecting CVEs in real world programs?

- **RQ3**: How useful is *EndWatch* in finding zero-day non-termination bugs?

**Tools under comparison.** We selected 6 state-of-the-art tools in the experiments: 2LS [27], AProVE [5], CPAchecker [3], UAutomizer [4], Dynamite [28], and Loopster [29]. Among them, UAutomizer, AProVE, 2LS, and CPAchecker achieved outstanding results in the past SV-Comp competitions. Dynamite is a termination and non-termination verification tool based on a dynamic approach. Loopster is a static approach which also computes path summaries.

In the experiments, we utilized the versions of 2LS, CPAchecker, and UAutomizer as supplied by the SV-COMP2023 competition, and utilized AProVE as supplied by the SV-COMP2022 competition. As for Dynamite, we employed the released Docker version available on GitHub [30]. Since Loopster is not available, we manually analyzed and identified the cases that could be handled by Loopster. We had intended to select VeriFuzz (FuzzNT) [31], a fuzzing-based tool, in our evaluation. However, it failed on almost all of our non-termination benchmarks, and as a result, we did not include its results in the paper.

Note that there are some other techniques [12]–[18] that can be used to detect performance issues. However, it is important to note that these methods are not suitable for comparison with non-termination detectors, as they are not designed to detect or confirm non-termination bugs.

**Existing benchmarks.** As the selected baselines are only capable of analyzing small programs, we selected benchmarks that contain curated programs with simplified program features. To this end, we choose the existing benchmarks that are widely recognized in termination analysis, such as those from SV-COMP2023 [19] and TermComp2022 [20]. The selected benchmarks encompass programs written in various languages, including C and Java. In our experiments, we only focused on C programs. The SV-COMP2023 benchmark comprises 40,604 C programs, grouped into 135 categories, while the TermComp2022 benchmark consists of 1,332 C programs classified into 11 categories. *EndWatch* is designed for non-termination loop detection, hence from the SV-COMP2023 benchmark, we selected categories labeled as "*loop*" and "*termination*". Using the "*false*" label, we further gathered 123 non-termination cases from its configuration file (.yml file). Similarly, we collected 130 "false-termination" benchmarks from TermComp2022, after removing 44 duplicated cases. We then excluded 15 non-termination cases resulting from infinite recursion, as the focus of our comparison is on non-termination loops. In total, we have 194 benchmark programs (112 from SV-COMP2023 and 82 from TermComp2022) in the experiment.

Furthermore, we also incorporate a new benchmark called OSS_Bench, which was recently released by [6]. The programs in this benchmark are extracted from real-world projects, making them more representative of real programs. Specifically, this benchmark was created to evaluate the capability of non-termination detection tools in handling real-world projects with special features. The benchmark comprises both non-termination and corresponding termination cases, which are divided into loop-caused and recursion-caused categories. Out of 118 programs, we only chose 44 loop-caused non-termination programs.

**CVE Programs.** To evaluate *EndWatch* on real-world programs, we gathered 20 CVEs related to infinite loop from the CVE website [32] for the past four years (2019-2022). The CVEs were filtered to only include those related to C/C++ programs and only the reproducible ones were considered. The collected CVEs came from 12 different projects including *libjpeg*, *Wireshark*, *Gpac*, *nasm*, *PDFResurrect*, *zziplib*, *picoquic*, *gdk-pixbuf*, *libsixel*, *cairo*, *Exiv2*, and *ProFTPD*.

The new programs are simplified based on the following principles: ❶ *Context Simplification* - The loops were simplified by identifying loop iterators and loop conditions, retaining relevant variables and data structures which may change the variables in the iterators, and removing instructions that do not affect loop termination. ❷ *Function Rewriting* - Functions affecting iterations were kept and others were removed. ❸ *Reserve Name and Types* - Consistent naming, typing of variables, and functions were maintained. ❹ *Making Benchmark Executable* - Following the benchmark [19], we add non-deterministic initialization of variables to make them executable.

**Real-world Projects.** Additionally, to evaluate the ability to detect zero-day bugs, we chose 5 high-star applications, *i.e.*, the OpenCV (66.1k stars) [33], Draco (5.5k stars) [34], ImageMagick (8.4k stars) [35], Libbpf (1.4k stars), [36], keystone (2k stars) [37].

All experiments are conducted in Ubuntu 22.04 system on a 3.9GHz 6-core AMD Rayzen processor with 16 GB RAM.

### A. Evaluation on Benchmark (RQ1)

*1) Setup:* For every case in the selected benchmark, we set a 900-second time limit, which is the same setting as the SV-COMP2023 competition. To minimize randomness, we repeated the testing process 5 times. We consider the non-deterministic variables (*e.g.*, `_VERIFIER_nondet_int()`) as the input variables for testing. In regards to 2LS, AProVE, CPAchecker, and UAutomizer, we adjusted their settings to match those of SV-COMP2023. For Dynamite, we utilized the option "`--nonterm`". The results such as "*timeout*", "*UNKNOWN*", "*MAYBE*", and "*TRUE/YES*" (representing termination) are considered not correct.

For the existing benchmarks with small programs, we use the online mode of *EndWatch*. Firstly, we perform static analysis to instrument the oracles (*i.e.*, SRC and revisit monitoring), and then we use AFL [11] to generate test cases for the instrumented programs. The total time taken by *EndWatch* includes both the time taken for static analysis and the time taken for testing.

*2) Result:* Table I presents the performance of the various tools on different benchmarks. In general, for benchmarks SV-COMP2023 and TermComp2022, it can be observed that

TABLE I: The results on benchmarks including SV-COMP2023, TermComp2022 and real-world OSS benchmark. *Total* represents the total number of programs in each category. # represents the number of correctly detected. The best value for each raw (benchmark) is highlighted in bold.

| Category | Total | *EndWatch* | | 2LS | | AProVE | | CPAchecker | | UAutomizer | | Dynamite | | *Loopster |
|---|---|---|---|---|---|---|---|---|---|---|---|---|---|---|
| | | # | Time(s) | # | Time(s) | # | Time(s) | # | Time(s) | # | Time(s) | # | Time(s) | # |
| loop-acceleration | 2 | **2** | **2.18** | 2 | 0.18 | 2 | 3.00 | 0 | 2.93 | 2 | 6.17 | 2 | 64.19 | 0 |
| loop-crafted | 1 | **1** | **0.41** | 1 | 6.26 | 1 | 1.79 | 0 | 0.70 | 1 | 2.47 | 0 | 1.36 | 0 |
| loop-invariants | 7 | **7** | **6.09** | 1 | 5400.06 | 1 | 5526.99 | 6 | 23.12 | 1 | 77.13 | 5 | 961.03 | 7 |
| loop-invgen | 2 | 1 | 900.74 | **1** | **0.13** | 0 | 902.93 | 0 | 1.39 | 0 | 3.62 | 0 | 2.80 | 1 |
| loop-lit | 1 | 0 | 901.32 | 0 | 710.98 | **1** | **2.15** | 0 | 855.81 | 0 | 2.15 | 1 | 23.47 | 0 |
| loops | 9 | **9** | 19.30 | 6 | **1.30** | 8 | 831.87 | 6 | 12.49 | 5 | 57.67 | 5 | 39.73 | 7 |
| termination-memory-alloca | 2 | **2** | **2.45** | 0 | 0.27 | 2 | 3.01 | 0 | 1.47 | 2 | 6.48 | 0 | 2.81 | 0 |
| termination-15 | 6 | **6** | **4.59** | 0 | 0.85 | 2 | 1909.09 | 0 | 4.38 | 6 | 31.65 | 0 | 5400.14 | 8 |
| termination-crafted | 11 | 10 | 906.53 | 9 | 905.47 | 9 | 1851.51 | 7 | 254.98 | **10** | **44.91** | 3 | 66.58 | 0 |
| termination-memory-linkedlists | 4 | **4** | 19.22 | 0 | **0.35** | 4 | 34.78 | 0 | 3.63 | 4 | 41.88 | 0 | 5.48 | 0 |
| termination-restricted-15 | 34 | **34** | **20.16** | 34 | 47.38 | 30 | 4272.86 | 25 | 3639.80 | 32 | 200.21 | 24 | 1324.87 | 13 |
| termination-nla | 19 | 15 | 3740.72 | 4 | 13501.31 | 0 | **22.55** | 3 | 3499.69 | 3 | 2234.76 | **17** | 1159.50 | 0 |
| termination-bwb | 14 | **14** | **9.76** | 11 | 2700.73 | 8 | 946.67 | 1 | 26.05 | 2 | 42.44 | 6 | 168.2 | 0 |
| AProVE_memory_alloca | 4 | **4** | **2.85** | 0 | 23.76 | 3 | 965.63 | 0 | 6.80 | **4** | 29.67 | 0 | 5.58 | 0 |
| SV-COMP_Mixed_Categories | 12 | 1 | 9917.56 | 2 | **1.76** | 3 | 9323.80 | **8** | 4256.97 | **8** | 3683.94 | 0 | 16.78 | 3 |
| SV-COMP_Termination_Category | 6 | **6** | 4.65 | 4 | 1800.36 | 4 | 3237.36 | 4 | **13.23** | **6** | 17.18 | **6** | 1369.29 | 3 |
| Ultimate | 16 | **16** | 21.09 | 11 | 3990.27 | 13 | 2122.94 | 10 | **161.76** | **16** | 54.11 | 9 | 1486.43 | 11 |
| Ton_Chanh_15 | 13 | 10 | 2719.04 | 3 | 3.07 | 8 | 2917.00 | **10** | **62.26** | 12 | 3706.93 | 6 | 233.69 | 11 |
| Stroeder_15 | 31 | **30** | 948.05 | 18 | 10533.43 | 26 | 6754.48 | 20 | **276.19** | 28 | 279.07 | 18 | 2009.19 | 26 |
| OSS_Bench | 44 | **36** | 7279.95 | 26 | 9574.82 | 21 | 9418.87 | 14 | **3040.24** | 18 | 7606.90 | 7 | 5455.60 | 4 |
| Total | 238 | **208** | 35566.27 | 133 | 49202.76 | 146 | 51049.26 | 114 | **16143.89** | 160 | 18129.36 | 109 | 19796.72 | 94 |

TABLE II: The number of programs correctly handled by the SRC (R1) and the revisit monitoring (R2), as well as the average time taken.

| Benchmark | Total | #C | | Avg Time(s) | | |
|---|---|---|---|---|---|---|
| | | R1 | R2 | R1 | R2 | Testing |
| SVComp | 105 | 54 | 51 | 1.25 | 0.48 | 3.15 |
| TermComp | 67 | 50 | 17 | 1.52 | 0.83 | 4.11 |
| OSS_Bench | 36 | 12 | 24 | 1.55 | 1.19 | 8.15 |
| Total Num& Avg Time | 208 | 116 | 92 | 1.38 | 0.73 | 4.43 |

TABLE III: The result on handling different features of benchmark programs. The best value for each row (feature) is highlighted in bold.

| Features (Total) | *EndWatch* | | 2LS | AP. | CPA. | UA. | Dyn. | Lo. |
|---|---|---|---|---|---|---|---|---|
| | R1 | R2 | | | | | | |
| Integer(165) | **101** | 34 | 95 | 100 | 97 | 111 | 85 | 94 |
| Array(20) | 5 | **14** | 8 | 9 | 0 | **14** | 2 | 0 |
| Pointer(16) | 0 | **16** | 0 | 10 | 0 | 12 | 0 | 0 |
| Bit calculation(20) | 6 | 14 | **15** | 0 | 4 | 4 | 5 | 0 |
| Data structure(6) | 0 | **6** | 1 | 5 | 0 | **6** | 0 | 0 |
| Function (20) | 4 | **16** | 8 | 4 | 10 | 12 | 4 | 0 |
| Overflow(9) | 6 | 3 | **8** | 0 | 3 | 2 | 1 | 0 |

*EndWatch* correctly detects 208 infinite loops, representing 87% of the total, which is more than the results of state-of-the-art tools. Specifically, 2LS correctly verified 133 loops (56%), AProVE correctly verified 146 loops (61%), CPAchecker correctly verified 114 loops (48%), UAutomizer correctly verified 160 loops (67%), Dynamite correctly verified 109 loops (46%) and Loopster can correctly verify 94 loops (39%).

Additionally, *EndWatch* has a 1.4× faster runtime compared to AProVE, but almost 2× slower than CPAchecker and UAutomizer. This is because if *EndWatch* encounters a case it cannot handle, it continues to run until the timeout, while the other tools may return *"UNKNOWN"* in a short time if they fail to verify a case. For instance, in categories such as *"termination-restricted-15"* is 180× faster than CPAchecker and 10× faster than UAutomizer, when it did not reach the timeout.

We also observed that Dynamite failed on many programs. Our in-depth analysis revealed two possible reasons: Firstly, Dynamite may not support some data types or structures. For instance, it returns "ERROR" when it encounters certain pointer or function calls. Secondly, Dynamite was unable to obtain the recurrent set for a lasso until it exceeded its maximum recursion depth.

The results of the real-world OSS benchmark (OSS_Bench) reveal that *EndWatch* is better at handling unpredictable conditions compared to the other tools. Out of the total cases, *EndWatch* correctly verified 82% (the highest percentage), while 2LS and AproVE verified 59% and 48% respectively, and CPAchecker and UAutomizer verified 32% and 41% respectively. On the other hand, Dynamite and Loopster, which performed poorly on this benchmark, correctly verified only 16% and 10%, respectively.

Table II provides detailed results of programs that are correctly handled by SRC (R1) and revisit monitoring (R2). The first and second columns represent the benchmark and total number of programs in each benchmark, respectively. The third column (#C) shows the number of programs correctly handled by SRC or revisit monitoring. The last column displays the average time taken by SRC, revisit monitoring, and fuzzing in each program. The results indicate that most of programs in SV-COMP and TermComp can be handled by SRC (60%). Revisit monitoring is more effective in handling real-world benchmarks, such as OSS_Bench (67%). The testing takes more time (4.43s) compared to static analysis (1.38s and 0.73s), indicating that the static analysis is lightweight.

We further show the capabilities of *EndWatch* and the other tools in handling different scenarios, such as those involving arrays, functions, pointers, etc. This was done by categorizing the programs into different features as shown in Table III. The first column indicates the feature of the program that causes non-termination of the loop, while the second column shows how many cases are handled by SRC (R1) and dynamic checking (R2), respectively.

TABLE IV: Results on CVE programs. R1 and R2 refer to the cases handled by SRC and revisit monitoring, respectively.

| Projects | CVEs | Org | Sim | | | | | | |
|---|---|---|---|---|---|---|---|---|---|
| | | EndWatch | EndWatch | 2LS | APr. | CPA. | AU. | Dyn. | Lo. |
| Libjpeg | CVE-2022-37768 | ✓(R2) | ✓(R2) | ✓ | U | U | ✗ | U | U |
| | CVE-2022-35166 | ✓(R2) | ✓(R2) | U | U | U | U | U | U |
| Wireshark | CVE-2022-0586 | ✓(R2) | ✓(R2) | ✓ | U | U | U | U | U |
| | CVE-2021-4184 | ✓(R2) | ✓(R2) | U | U | U | U | U | U |
| | CVE-2021-4185 | ✓(R2) | ✓(R2) | U | U | U | U | U | U |
| | CVE-2020-26575 | ✓(R2) | ✓(R2) | U | U | U | U | U | U |
| | CVE-2019-16319 | ✓(R2) | ✓(R2) | U | U | U | U | U | U |
| | CVE-2019-10897 | ✓(R2) | ✓(R2) | U | U | U | U | U | U |
| Gpac | CVE-2021-45297 | ✓(R1) | ✓(R1) | ✗ | U | ✓ | U | U | U |
| | CVE-2021-44924 | - | - | - | - | - | - | - | - |
| | CVE-2021-40592 | - | - | - | - | - | - | - | - |
| nasm | CVE-2021-45257 | ✓(R2) | ✓(R2) | ✗ | U | U | ✓ | U | U |
| PDFResurrect | CVE-2021-3508 | ✓(R2) | ✓(R2) | ✗ | U | U | U | U | U |
| zziplib | CVE-2020-18442 | ✓(R2) | ✓(R2) | ✗ | U | U | U | U | U |
| picoquic | CVE-2020-24944 | ✓(R2) | ✓(R2) | ✗ | U | U | U | U | U |
| gdk-pixbuf | CVE-2020-29385 | ✓(R2) | ✓(R2) | ✓ | U | U | ✓ | U | U |
| libsixel | CVE-2019-3573 | ✓(R1) | ✓(R1) | ✗ | U | U | U | U | U |
| cairo | CVE-2019-6462 | ✓(R1) | ✓(R1) | U | U | U | U | U | U |
| Exiv2 | CVE-2019-20421 | ✓(R2) | ✓(R2) | U | U | U | U | U | U |
| ProFTPD | CVE-2019-18217 | ✓(R1) | ✓(R1) | U | U | U | U | U | U |

TABLE V: Hangs and non-termination bugs found

| Projects | #Hangs | Non-termination |
|---|---|---|
| OpenCV | 1,624 | 1 |
| Keystone | 2,310 | 1 |
| Draco | 2,241 | 1 |
| ImageMagick | 649 | 0 |
| Libbpf | 363 | 1 |

Among these results, *EndWatch* can detect 61% of integer cases through R1, but requires revisit monitoring (R2) to handle features like Pointer, Bit calculation, Data structure, and Function that are hard to be analyzed statically. *EndWatch* outperforms the other tools in handling non-integer cases. For example, UAutomizer performs best among these tools for programs involving functions, handling 52% of cases, while *EndWatch* correctly detects 87% of cases. It is because *EndWatch* benefits from the dynamic fuzzing strategy that can ignore complex features and provide accurate context information, which is more advantageous than the existing static analysis-based methods.

> **Answer to RQ1:** *EndWatch* outperforms all existing approaches on existing benchmarks. With *EndWatch*, we can correctly detect 87% non-termination benchmarks. Due to the state revisit monitoring, *EndWatch* shows a significant performance advantage at handling loops with special features such as array, pointer, bit calculation, data structure, and function.

### B. Evaluation on CVE Programs (RQ2)

*1) Setup:* We conducted two experiments to answer this research question. Firstly, we applied *EndWatch* to the collected real-world projects to demonstrate its capability in real-world scenarios. *EndWatch* was configured in offline mode for these large projects, *i.e.*, whether our oracles can identify non-termination when provided with the non-terminating inputs. Secondly, to compare with the baselines that cannot directly handle large programs, we also evaluated all of the tools on simplified versions of these programs within 900 seconds.

*2) Result:* Table IV shows the results on CVE programs. *Org* refers to the results on the original project while *Sim* refers to the results on the simplified versions. In the table, "✓" represents a correct verification, "✗" means a non-termination bug is incorrectly verified as "termination", and "U" demonstrates that the tool is unable to handle this case within the specified time (900s), *e.g.*, "UNKNOWN", "TIMEOUT", and "MAYBE". The results on the collected CVEs show that the oracles provided by *EndWatch* are able to detect the non-termination effectively. We observed that 4 non-termination bugs were detected by SRC while 14 bugs were detected by our revisit monitoring. This could be due to the fact that more non-termination bugs occur in non-linear loops, which often involve pointers, data structures, arrays, and functions, and are more prone to non-termination due to their complexity and potential for subtle bugs. *EndWatch* failed in CVE-2021-44924 and CVE-2021-40592. The reason is that the infinite loop in both CVE-2021-44924 and CVE-2021-40592 exists within a dynamic shared library, *i.e.*, the files end with ".so", which cannot be instrumented and analyzed by *EndWatch*.

Compared to the simplified programs, we can observe that *EndWatch* is more effective in detecting most of the CVEs. Our in-depth analysis reveals that AProVE fails in all the programs due to its limitation to support data structures, which are widely used in these programs. Dynamite and Loopster failed because they cannot handle functions and pointers well. On the other hand, 2LS only successfully handles four cases, but also produces five incorrect results.

> **Answer to RQ2:** *EndWatch* is highly effective in detecting non-termination bugs in real-world programs. Our experiments demonstrate that *EndWatch* can correctly handle 90% of the non-termination bugs in the original CVE programs, outperforming state-of-the-art tools such as 2LS, AProVE, CPAchecker, UAutomizer, Dynamite, and Loopster. These tools were only able to detect 15%, 0%, 5%, 10%, 0%, and 0% of non-termination bugs in the simplified versions of the CVE programs, respectively.

### C. Zero-Day Bug Detection (RQ3)

*1) Setup:* We configured *EndWatch* in offline mode to detect non-termination. Specifically, we first run AFL to generate test cases and collect the *hang* test cases. Each project was fuzzed with 240 hours. Then we instrumented the programs with *EndWatch* and fed the hangs to the instrumented programs to identify the non-termination inputs.

*2) Result:* Four infinite loop bugs have been found, and two of them have been confirmed and fixed. Among them, 3 bugs are detected by Revisit Monitoring and 1 bug is found by SRC. Table V shows the number of hangs found by fuzzing and non-termination detected by *EndWatch*.

Another possible approach to identify non-termination bugs is to manually analyze long-running tests (hangs) collected from existing tools. However, this method has some chal-

```
1 ...
2 while (true){
3      IdentityOpsMap::iterator
4      nextIt = identity_ops.find( it->second );
5      if (nextIt != identity_ops.end())
6           it = nextIt;
7      else
8           break;
9 }
```

Fig. 4: Issue #22709 in OpenCV.

lenges. Firstly, it can be time-consuming, inaccurate, and requires significant human effort. Based on our experience, it takes tens of minutes or even hours to analyze and confirm a hang test case, especially when the project logic is complex. Secondly, it is not trivial to set a suitable time threshold for collecting hangs, as it depends on the project. A larger threshold may reduce some hangs, but it requires more time to filter, especially when there are too many hangs. For example, OSS-Fuzz [38] can detect a large number of hangs, but few non-termination bugs are reported from them. Therefore, *EndWatch* is designed to automatically detect and confirm non-termination bugs.

Figure 4 shows a new non-termination bug we discovered in OpenCV. This bug is due to the cyclic assignment in the map data structure. Specifically, the map iterator is utilized to search for a corresponding value using a key "it->second". However, when the key is equal to the value in the map iterator, "identity_ops.find(it->second)" returns the same iterator, making "it" remains unchanged. This issue was difficult to detect using existing tools due to the complexity of the data structures and pointers involved as well as the complex context in the whole project. However, by utilizing revisit monitoring, *EndWatch* was able to record the relevant variables as states (*i.e.*, "it") and check for its revisits during testing, leading to the detection of the infinite loop. More detail and case studies can be found on our website [25].

**Answer to RQ3:** *EndWatch* is useful in detecting new non-termination bugs in real-world projects.

## VI. THREATS TO VALIDITY

The selected benchmarks could be a threat to validity. To mitigate this threat, we select multiple benchmarks including the standard benchmarks, the OSS benchmarks, the projects with CVEs, and other projects. Another threat comes from the randomness in our fuzzing testing. Because mutation in fuzzing is a heuristic method, which may affect the generalizability of the RQ1 result. To alleviate this issue, we repeated 5 times and calculate the average time. The step interval and the corresponding hyper-parameter in Section IV-C could be a threat to validity. Different parameters setting may affect the experiment result. The setting in the paper may miss some non-termination if we cannot observe the revisit.

## VII. RELATED WORK

### A. Static Analysis for Non-termination

Non-termination is a classic topic that has been studied for a long time. Many static researches ( [39], [21], [40], [41], [42]) aim to find a recurrent set. Zhang [39] proposed a path-based approach for identifying non-termination in simple loops (rather than nested loops). He establishes the impracticality of certain paths within a loop in a symbolic manner, enabling the verification of non-termination. Gupta *et al.* [21] introduced a concept named lasso. They enumerated a lasso-shaped path for counter example, and then to determine the feasibility, they searched for a recurrent set by exploring the program symbolically. Chen *et al.* [40] extended the recurrent set to a closed recurrence set and reduced it to an under-approximation nonlinear program. Cook *et al.* [41] intruded live abstraction to make over-approximation on non-linear programs for recurrent set inferring. Larraz *et al.* [42] used Max-SMT to generate quasi-invariant, and take it as a property to demonstrate the non-termination. Our work is also interested in recurrent set detection. However, we use the test oracle to identify whether the execution has fallen into a recurrent set, which is a subset of the non-termination input, rather than to make a static verification.

Except for recurrent set detection, some researchers focus on infinite state verification. Le *et al.* [43] adopted a Hoare-style forward verification and incorporated unknown pre/post-pre-dicate to discover both termination and non-termination properties. Frohn *et al.* [44] proposed the approach named loop acceleration. They accelerated the terminating loops to prove the reachability of non-terminating configurations. Different from the non-termination checking, Xie *et al.* [29] adopted path dependency automation, and inferred the termination and non-termination of a single path and interleaving path via monotonicity analysis. *EndWatch* differs from Loopster in three aspects: 1) Loopster can only detect non-termination caused by single stuck paths while *EndWatch* can detect non-termination caused by interleaving of multiple paths. 2) Loopster determines the termination or non-termination by checking satisfiability of the whole program, while *EndWatch* calculates explicit SRCs of individual loops, which are used as oracles for testing. 3) Loopster is pure static and inapplicable in large programs, while *EndWatch* can handle real-world program, especially due to state revisit checking.

### B. Dynamic Analysis for Non-termination

Compared with static analysis, dynamic analysis for non-termination is relatively more similar to our work. Le *et al.* [28] developed Dynamite which can verify termination and non-termination for a loop by dynamic analysis. They first dynamically execute the program to acquire dynamic snapshots. Then they use these snapshots and a template to infer a ranking function for termination verification or learn a recurrent set for non-termination verification. Karmarkar *et al.* [31] proposed FuzzNT for non-termination testing. They used a guess-and-check approach to guess a prefix by AFL and then they check non-termination by abstract interpretation-based analysis. Similar to them, we also run the program dynamically, but we verify the non-termination by test oracle rather than to "guess" a candidate result, which is more scalable in real world programs.

*C. Detecting performance issues*

Many approaches have been proposed to explore effective ways on low-performance fuzzing. [12]–[16]. SlowFuzz [12] adopts the number of executed instructions as the fitness in genetic algorithm for a worse performance. Perfuzz [16] further takes the visited paths into account and uses feedback-directed mutational fuzzing to help generate slowdown input. HotFuzz [13] constructs a test harness for every function, and takes the number of executed instructions as fitness in their generic algorithm to slow down the execution. Different from the former feedback-based method, Singularity [15] conducts a black box fuzzing, and finds slowdown cases by generating the input with a specific pattern. ReScue [14] adopts a backtracking search approach to find the target string which can induce Regex Dos. Some of the research is mainly focused on loop efficiency [17], [18]. CARAMEL [17] adopts the static analysis to find the variable that makes a loop executes redundant iterations. Dhok *et al.* [18] detected performance issues by finding redundant functions. Detecting performance issues is different from proving non-termination, as discussed in Section III-A.

VIII. CONCLUSION

In this work, we introduced *EndWatch*, a practical method for detecting non-termination in real-world programs. *EndWatch* combines static and dynamic analyses, where static analysis calculates non-termination oracles, and dynamic analysis generates test cases to violate these oracles. *EndWatch* can be applied in both offline and online modes, allowing it to handle programs of different sizes and complexities. We evaluated it on standard benchmarks and real-world projects. The results showed its effectiveness and usefulness, especially in detecting non-termination in real-world projects.

ACKNOWLEDGMENTS

This work was partially supported by the National Key R&D Project (2021YFF1201102), the National Natural Science Foundation of China (Grant No. 61872262), the Ministry of Education, Singapore under its Academic Research Fund Tier 1 (21-SIS-SMU-033), and NTU Start-Up Grant.